\documentclass[12pt]{iopart}

\usepackage{iopams}  
\usepackage{graphicx}
\usepackage{amssymb}
\usepackage{amsgen}
\usepackage{psfrag}

\begin{document}

\title {Landau levels of cold atoms in non-Abelian gauge fields}

\author{A. Jacob$^1$, P. \"Ohberg$^2$, G. Juzeli\={u}nas$^3$ and L. Santos$^1$}

\address{$^1$Institut f\"ur Theoretische Physik, Leibniz Universit\"at
Hannover, Appelstr. 2, D-30167, Hannover, Germany}
\address{$^2$ SUPA, School of Engineering and Physical Sciences, 
Heriot-Watt University, Edinburgh EH14 4AS, United Kingdom}
\address{$^3$Institute of Theoretical Physics and Astronomy of Vilnius 
University, A. Go\v{s}tauto 12, 01108 Vilnius, Lithuania}
\ead{andreas@itp.uni-hannover.de}
\begin{abstract}
The Landau levels of cold atomic gases in non-Abelian gauge 
fields are analyzed. In particular we identify effects on the energy spectrum 
and density distribution which are purely due to the non-Abelian character 
of the fields. We investigate in detail non-Abelian generalizations of both the 
Landau and the symmetric gauge. Finally, we discuss how these non-Abelian 
Landau and symmetric gauges may be generated by means of realistically 
feasible lasers in a tripod scheme.
\end{abstract}

\maketitle

\section{Introduction}
\label{sec:1}


Gauge potentials are crucial for the understanding of fundamental 
forces between subatomic particles. A simple example of a gauge potential 
is provided by the vector potential in the theory of electromagnetism \cite{Jackson}. 
In this example the different vector components are scalars, and hence 
they commute with each other. If the vector components of the gauge field are 
not scalars, but instead $N\times N$ matrices, with $N>1$, then 
it is in principle possible to have a situation where the different 
vector components do not commute. However, non-Abelian gauge fields are scarce in 
nature. Candidates so far have mainly been restricted to molecular systems \cite{mead1992}
which are largely approachable only through spectroscopic means. Other systems 
are liquid crystals which show the required non-Abelian symmetries
\cite{liquid}.


Experiments on cold quantum gases have reached an unprecedented degree of control,  
offering thus extraordinary possibilities for the analysis of the effects of gauge 
fields on atomic systems. A simple way of generating a gauge field in ultracold gases 
is obtained by rotating Bose-Einstein condensates with an angular frequency 
$\Omega \hat z$ (where we employ cylindrical coordinates $\{ \rho,\varphi,z\} $). 
In the corresponding rotating frame the Hamiltonian describing the rotating system
becomes one of a system subject to a symmetric 
Gauge field $\vec A=-m\Omega\rho\hat\varphi$, where $m$ is the atomic mass \cite{BookStringari}. 
Thus, a rotating condensate resembles a gas under the influence of a constant magnetic 
field $B_0=m\Omega$, and as a consequence many interesting phenomena, including 
e.g. Landau level physics and Quantum-Hall-like phenomena have been studied in rotating 
quantum gases \cite{Rotating1,Rotating2,Rotating3,Rotating4}. 


Due to their internal structure, ultracold atoms 
offer as well the possibility of creating non-Abelian gauge fields. 
A surprising and astoundingly elegant derivation and description of 
the emergence of non-Abelian gauge potentials 
was presented by Wilczek and Zee \cite{wilczek1984}. It was 
shown by these authors that in the presence of a general adiabatic motion of 
a quantum system with degenerate states, gauge potentials will appear 
which are traditionally only encountered in high energy physics to 
describe the interactions between elementary particles. 
Ultracold atomic clouds are particularly promising candidates 
for realising such scenarios, since the  
access to physical parameters is, from an experimental point of view, 
unprecedented. Extending the ideas of Wilczek and Zee, it was 
recently proposed that properly tailored laser beams coupled to degenerate 
internal electronic states can be employed to induce Abelian as well as non-Abelian 
gauge fields in cold-atom experiments 
\cite{juzeliunas2004,ruseckas2005,Juzeliunas2007}.
Alternatively, such gauge potentials can be constructed in an optical lattice 
using laser assisted state sensitive tunneling \cite{Jaksch,osterloh2005, Mueller,Demler}.


With the implementation of these proposals, ultracold atoms would offer 
a unique testbed for the analysis of non-trivial effects on 
the properties of multicomponent cold atomic systems in the presence of 
non-Abelian gauge fields. To the best of our knowledge 
these effects have been scarcely studied in the literature \cite{ruseckas2005,
Juzeliunas2007,osterloh2005,Clark}.
This paper is devoted to the analysis of 
non-Abelian effects on the spectral properties of ultracold atomic systems. 
In particular, we show how purely non-Abelian effects lead to the eventual destruction of the 
Landau level structure, and may significantly modify the ground state density profile of ideal 
quantum gases.


The structure of the paper is as follows.
In Sec.~\ref{sec:2} we study the generation of different forms of non-Abelian gauge fields, 
including non-Abelian constant fields, as well as the non-Abelian generalization of 
the Landau gauge. Sec.~\ref{sec:3} is devoted to the analysis of constant non-Abelian gauge
fields. Sec.~\ref{sec:4} discusses the non-Abelian Landau gauge, and in particular the destruction of 
the Landau level structure and the corresponding modified de Haas-van Alphen effect. In Sec.~\ref{sec:5} 
we discuss the non-Abelian symmetric gauge. Finally in Sec.~\ref{sec:6} we conclude and discuss some 
future promising directions.

\section{Laser-induced non-Abelian Gauge fields}
\label{sec:2}

In this section we discuss the generation of (possibly non-Abelian) gauge fields as those 
discussed in the following sections. There are at least two alternatives 
for the creation of non-Abelian gauge fields. One consists in employing 
two-component atoms in state-dependent optical lattices in the presence 
of appropriate laser arrangements \cite{Jaksch,osterloh2005}. 
A second possibility, which we 
shall explore in this paper, was recently proposed in 
Ref.~\cite{ruseckas2005}. 

In this second alternative, a non-Abelian gauge 
potential is constructed for atoms with a tripod electronic structure \cite{unanyan98,unanyan99}. 
Three lasers with appropriate polarizations couple 
the excited electronic state $|0\rangle$ and the groundstates $|j=1,2,3\rangle$, with  
corresponding Rabi frequencies $\Omega_j(\vec r)$ which parametrize
in the form:
$\Omega_1  = \Omega\, \sin\theta\, \cos\phi\, \mathrm{e}^{iS_1}, 
\Omega_2  = \Omega\, \sin\theta\, \sin\phi\,  \mathrm{e}^{iS_2},
\Omega_3  = \Omega\, \cos\theta\, \mathrm{e}^{iS_3}$.
For a fixed position $\vec{r}$ the Hamiltonian describing the laser-atom 
interaction may be diagonalized to give a set of dressed states.  
Under appropriate conditions two dressed states, so called dark states become decoupled from the other states:
\begin{eqnarray}
|D_1\rangle & = &\sin\phi \mathrm{e}^{iS_{31}}|1\rangle
-\cos\phi e^{iS_{32}}|2\rangle,
\label{eq:D1} \\
|D_2\rangle & = &\cos\theta \cos\phi e^{iS_{31}}|1\rangle
+\cos\theta\sin\phi e^{iS_{32}}|2\rangle  -\sin\theta |3\rangle ,
\label{eq:D2}
\end{eqnarray}
with $S_{ij}=S_i-S_j$. The dark states have zero eigenvalues and are separated by the energy $\hbar\Omega$ from the remaining eigenstates. The state of the atom can therefore be 
expanded in the dark state basis as  
$|\Phi\rangle =\Psi _{1}(\vec{r}) \left |
D_1(\vec{r})\right\rangle+\Psi_{2}(\vec{r}) \left|
D_2(\vec{r})\right\rangle$. The two-component spinor  $\vec\Psi=\{\Psi_1,\Psi_2\}^T$ obeys
a spinor Schr\"odinger equation of the form
\begin{equation}
i\hbar\frac{\partial}{\partial t}\vec\Psi=\left[ \frac{1}{2m}
(-i\hbar\vec\nabla -\hat{A})^2 + \hat V +\hat\Phi \right]\vec\Psi.
\label{eq:SE-reduced}
\end{equation} 
In this equation we observe the appearance of a 
$2\times 2$ vector potential of the form
\begin{eqnarray}
\hat{A}_{11} &=& \hbar\left(\cos^2\phi\vec\nabla S_{23}
+ \sin^2\phi\vec\nabla S_{13}\right)\, ,\nonumber \\
\hat{A}_{12} &=& \hbar\cos\theta\left(\frac{1}{2}\sin(2\phi)
\vec\nabla S_{12}-i\vec\nabla\phi\right)\, , \label{eq:A-special} \\
\hat{A}_{22} &=&\hbar\cos^2\theta\left(\cos^2\phi
\vec\nabla S_{13}+\sin^2\phi\vec\nabla S_{23}\right).\nonumber
\end{eqnarray}
The systems also presents a scalar potential of the form 
$\Phi_{ij}=\frac{\hbar^2}{2m}\vec{\kappa}_i^*\cdot\vec{\kappa}_j$, 
where 
\begin{eqnarray}
\vec {\kappa}_{1} &= \sin\theta \left(\frac{1}{2}\sin(2\phi)\vec\nabla S_{12}
+i\vec\nabla\phi\right),\\
\vec{\kappa}_{2} &=\frac{1}{2}\sin(2\theta)(\cos^2\phi\vec\nabla S_{13}
+\sin^2\phi\vec\nabla S_{23})-i\vec\nabla\theta.
\end{eqnarray} 
Finally, if the original states $|j=1,2,3\rangle$ experience an 
external potential $U_j(\vec{r})$, then 
\begin{eqnarray}
V_{11}&=&U_2\cos^2\phi+U_1\sin^2\phi, \\
V_{12}&=&\frac{U_1-U_2}{2}\cos\theta\sin(2\phi), \\
V_{22}&=&(U_1\cos^2\phi+U_2\sin^2\phi)\cos^2\theta+U_3\sin^2\theta. 
\end{eqnarray}
Recent advances in shaping 
both the phase and the intensity of light beams make it possible 
to achieve a remarkable versatility in controlling the gauge fields \cite{Control1,Control2}, 
provided the corresponding light fields obey Maxwell's equations. 

In the following we shall assume that the atoms are 
strongly trapped in the $z$-direction, hence they are confined to the $xy$ plane. Given two 
orthogonal vectors $\vec{\xi}$ and $\vec{\eta}$ 
on the $xy$ plane, 
we shall be interested in non-Abelian situations, in which 
$\hat {A}_{\xi}\equiv \hat{A}\cdot\vec{\xi}$
and $\hat {A}_{\eta}\equiv \hat{A}\cdot\vec{\eta}$, 
fulfill $[\hat{A}_\xi,\hat{A}_\eta]\neq 0$. 
This condition demands 
$(\vec{u}\times\vec\nabla\phi)_z \neq 0$ and/or 
$(\vec{u} \times\vec\nabla S_{12})_z \neq 0$, and/or
$(\vec\nabla S_{12} \times \vec\nabla\phi)_z \neq 0$, 
with $\vec{u}=(\cos^2\phi-\cos^2\theta\sin^2\theta)\vec\nabla S_{23} +
(\sin^2\phi-\cos^2\theta\cos^2\phi)\vec\nabla S_{13}$.

\subsection{Constant intensities}

We will consider first homogeneous intensity profiles, i.e. both $\phi$ and 
$\theta$ are now space independent. We choose the particular case with 
$\phi=\theta=\pi/4$. For constant $\phi$ 
the non-Abelian character demands 
$\vec\nabla S_{23}\times \vec\nabla S_{13}\neq 0$.
A simple laser arrangement fulfilling this condition is 
$S_{j3}=\alpha_j x + \beta_j y$, where 
$\alpha_j,\beta_j$ are constants such that  
$\alpha_2\beta_1\neq \alpha_1\beta_2$. 
The corresponding $x$ and $y$ components of the 
vector potential are of the form
\begin{eqnarray}
\hat{A}_x=\frac{1}{8}(\alpha_1+\alpha_2)(3\hat{1}+\hat\sigma_z)+
\frac{1}{2\sqrt{2}}(\alpha_1-\alpha_2)\hat\sigma_x, \\
\hat{A}_y=\frac{1}{8}(\beta_1+\beta_2)(3\hat{1}+\hat\sigma_z)+
\frac{1}{2\sqrt{2}}(\beta_1-\beta_2)\hat\sigma_x.
\end{eqnarray}
On the other hand, by choosing $V_j(\vec r)=\Delta E_j + U(\vec{r})$,
with 
$\Delta E_1=-(\hbar^2/16 m) 
[(\alpha_1^2-\alpha_2^2)+(\beta_1^2-\beta_2^2)]=-\Delta E_2$, and 
$\Delta E_3=-(\hbar^2/16 m) 
[(\alpha_1^2+\alpha_2^2)+(\beta_1^2+\beta_2^2)]$, one can prove 
that (up to an irrelevant constant) $\hat{V}+\hat{\phi}=U(\vec{r})$
with $U(\vec{r})$ a common trapping potential for all components.  

A gauge transformation eliminates the terms proportional to the 
identity matrix in $\hat A_x$ and $\hat A_y$. 
Let $\hbar \kappa_y=(\beta_1-\beta_2)/2\sqrt{2}$, 
$\hbar q_y=(\beta_1+\beta_2)/8$,  
$\hbar \kappa_x=(\alpha_1-\alpha_2)/2\sqrt{2}$, and 
$\hbar q_y=(\alpha_1+\alpha_2)/8$. 
A rotation 
$\hat\sigma_x\rightarrow\cos\eta \hat\sigma_x +\sin\eta \hat\sigma_z$, 
$\hat\sigma_x\rightarrow -\sin\eta \hat\sigma_x +\cos\eta \hat\sigma_z$, 
with $\tan 2\eta=\kappa_y/q_y$, 
provides $\hat A_y=\hbar \tilde q_y \hat\sigma_z$, with 
$\tilde q_y=\cos 2\phi q_y +\sin 2\phi \kappa_y$, and 
$\hat A_x=\hbar\tilde \kappa_x \hat\sigma_x +
\hbar \tilde q_x \hat\sigma_z$, with 
$\tilde \kappa_x=(\cos2\phi \kappa_x-\sin 2\phi q_x)$, and 
$\tilde q_x=(\cos2\phi q_x+\sin 2\phi \kappa_x)$. 
Hence, we recover exactly the same form which was discussed in Sec.~\ref{sec:3}.

\subsection{Landau-like gauge}

In this subsection we shall consider the case $S_{13}=S_{23}=S$. In that 
case the non-Abelian character demands $(\vec\nabla S \times \vec\nabla\phi)_z\neq 0$.
We will choose the phase $S=\kappa x$, and $\phi=qy$, which gives a non-Abelian gauge potential
unless $\kappa=0$ or $q=0$. In addition we take $\cos\theta = x/R_c$, where $R_c^2=x^2+(z-z_c)^2$, such that for the relevant $x$-range, $|x|<<z_c$ is fulfilled. As a consequence, and up to 
first order in $(x/z_c)$ we obtain:
\begin{equation}
\hat{A}\simeq \hbar \kappa (\hat{1}+\hat\sigma_z) \hat x
+B_0 x \hat\sigma_y \hat y.
\end{equation}
where $B_0=\frac{q}{z_c}$. Note that although $x\ll z_c$, 
$B_0$ can actually have large values. 
In addition, and again up to first order in $(x/z_c)$, 
we obtain $\hat{V}+\hat{\phi}=U(\vec{r})$, 
if $V_1(\vec{r})=V_2(\vec{r})=\hbar^2 q^2/2m+U(\vec{r})$, and 
$V_3(\vec{r})=\hbar^2/2mz_c^2$. Using a simple gauge transformation  
$\Psi\rightarrow \exp{i\kappa x}\Psi$ to eliminate the 
identity matrix term in $\hat{A}_x$, and 
applying a unitary spin transformation $U^\dag\hat{A}U$, with 
$U=(\hat\sigma_z+\hat\sigma_y)/\sqrt{2}$, we obtain 
$\hat{A}\simeq \hbar \kappa \hat\sigma_y \hat x
+B_0 x \hat\sigma_z \hat y$, which is indeed exactly 
the same Landau-like gauge that we employ in Sec.~\ref{sec:4}.
A simple laser arrangement which would lead to this 
particular gauge is provided by 
\begin{eqnarray}
\Omega_1&=&\Omega\cos qy e^{i\kappa(x+y+z)/2}, \\
\Omega_2&=&\Omega\sin qy e^{i\kappa(x+y+z)/2}, \\
\Omega_3&=&\Omega\frac{x}{z_c} e^{i\kappa(x-y+z)/2},
\end{eqnarray}
where we assume the illuminated atoms are confined to a region for which $|x|\ll z_c$ holds.

\section{Constant non-Abelian gauge}
\label{sec:3}

Let us consider a constant matrix gauge of the form 
$\hat A=(\hat A_x,\hat A_y,0)$. We have already shown that these 
fields can be generated in a tripod scheme using a simple 
laser arrangement. Then, the Hamiltonian of the 2D system becomes:
\begin{equation}
\hat H=\frac{1}{2m}
\left [ (\hat p_x+\hat A_x)^2+(\hat p_y+\hat A_y)^2\right ]. 
\end{equation}
In the Abelian case $[\hat A_x,\hat A_y]=0$. 
We can therefore choose a common eigenbasis for both matrices:
$\hat A_x/\hbar={\rm diag}\{q_{1x},q_{2x}\}$, 
$\hat A_y/\hbar={\rm diag}\{q_{1y},q_{2y}\}$,
As a consequence, we recover two independently 
displaced quadratic spectra 
$E_j(\vec k)=\frac{\hbar^2}{2m} 
\left (\vec k+\vec q \right )^2$, where 
$\vec q_j=(q_{jx},q_{jy})$.

In the non-Abelian case, on the other hand, we 
cannot simultaneously diagonalize both matrices, and as a 
consequence the spectrum becomes distorted.
Let us consider a simple, but representative, case, namely 
$\hat A_x=q_x\hat\sigma_x$, $\hat A_y=q_y\hat\sigma_z$.
Employing the Fourier-like transformation
\begin{equation}
\vec\psi(x,y)=\sum_{kx,ky} e^{ik_y y \hat\sigma_z}
\left (\frac{1+i\hat\sigma_y}{\sqrt{2}}\right )
e^{ik_x x\hat\sigma_z}\vec\phi(k_x,k_y) 
\end{equation}
with $k_{x,y}=2\pi n_{x,y}/L$, 
we may transform the time-independent Schr\"odinger equation 
$E\vec\psi(x,y)=\hat H\vec\psi(x,y)$ into 
\begin{equation}
\frac{2mE}{\hbar^2}\vec\phi(k_x,k_y)=
\left [k_x^2+q_x^2+(k_y+q_y)^2 \right ]\vec\phi(k_x,k_y) +2q_xk_x\vec\phi(k_x,-k_y).
\end{equation}
Diagonalizing the system of equations for $\phi(k_x,\pm k_y)$, 
we obtain two eigenenergies
\begin{equation}
\frac{2mE_\pm}{\hbar^2}=
 k_x^2+q_x^2+k_y^2+q_y^2\pm 2\sqrt{k_x^2q_x^2+k_y^2q_y^2} 
\end{equation}
Note that in the Abelian situation $q_x=0$ (or $q_y=0$), and, as expected 
there is no coupling between momenta in different directions. 
However, due to the non-Abelian character, even for a
constant gauge there is a non trivial coupling between the 
different directions.

\section{Landau-like Non-Abelian gauge}
\label{sec:4}

\subsection{Periodic boundary conditions}

We consider in the following a matrix 
generalization of the Landau gauge, 
namely $\hat A=(\hbar\kappa \hat M_x,B_0\hat M_y x,0)$ 
(the usual Landau gauge is of the form $(0,B_0x,0)$). 
We will assume that the matrices $\hat M_x$ and $\hat M_y$ are constant.
Then the Hamiltonian of the 2D system becomes:
\begin{equation}
\hat H=\frac{1}{2m}
\left [ (\hat p_x+\hbar\kappa \hat M_x)^2+(\hat p_y+B_0\hat M_y x)^2\right ]. 
\end{equation}

We first discuss the typical text book situation, in which the 
particles (which are assumed to be confined on the $xy$-plane) 
are considered as confined in a 2D box of side $L$ 
with periodic boundary conditions (i.e. a toroidal configuration).  
We are particularly interested in how the 
non-Abelian character of the fields destroys the 
usual Landau-level structure of the energy eigenstates.  
In the following subsection we shall discuss 
a slightly different scenario closer to actual 
experimental conditions.

As in Sec.~\ref{sec:3}, if $[\hat M_x,\hat M_y]=0$, one 
can find a common eigenbasis $\{\vec e_1,\vec e_2\}$, such that in this basis 
$\hat M_x={\rm diag}\{\gamma_1,\gamma_2 \}$, and 
$\hat M_y={\rm diag}\{\lambda_1,\lambda_2 \}$, and hence the 
Hamiltonian is also diagonal in this basis. 
Since we assume periodic boundary conditions 
we can thus consider wavefunctions of the form
\begin{equation}
\vec\psi_j(\vec r)=\sum_{n_y}e^{i\frac{2\pi}{L}n_y y+
i\kappa\gamma_j q}v_j(n_y,x)\vec e_j,
\end{equation}
such that
\begin{equation}
E v_j(q)=
\left [\frac{\hat p^2}{2m}+\frac{1}{2}m\omega_j^2 q^2\right ]v_j(q).
\end{equation}
where $q=x+\frac{2\pi\hbar n_y}{LB_0\lambda_j}$, $p=-i\hbar\partial/\partial q$, and 
$\omega_j=B_0|\lambda_j|/m$ is the cyclotron frequency for the state $j$.
Hence, for the Abelian case we obtain two different sets of Landau levels with energies 
$E_j(n)=\hbar\omega_j(n+1/2)$, and degeneracies $g_j=B_0\lambda_j L^2/2\pi\hbar$.
Note that if $|\lambda_1|=|\lambda_2|$, as it is the case for $M_y=\hat\sigma_z$, 
then the two sets of Landau levels are degenerate.

Let us now discuss what happens if on the contrary $[\hat M_x,\hat M_y]\ne 0$. 
We work (without lack of generality) in the basis in which 
$M_y=\hat\sigma_z$. Note that the Ansatz 
\begin{equation}
\vec\psi(\vec r)=\sum_{n_y}e^{i\frac{2\pi}{L}n_y y \hat\sigma_z}\vec u(n_y,x),
\end{equation}
also fulfills periodic boundary conditions. 
We insert this Ansatz in the eigenvalue equation to 
obtain
\begin{eqnarray}
E\vec u(n_y,x)&=&\left [ \frac{\hat\Pi^2}{2m}+ 
\frac{\hbar^2}{2m}\left ( \frac{2\pi n_y}{L}+\frac{B_0}{\hbar} x \right )^2 
\right ] \vec u(n_y,x) \nonumber \\ 
&+&
\left [
\hat\sigma_z [ \hat\Pi^2, \hat\sigma_z] \right ]\left [
\frac{\vec u(n_y,x)-\vec u(-n_y,x)}{4m}
\right ].
\end{eqnarray}
where $\hat\Pi=\hat p_x+\hbar\kappa\hat M_x$. For the Abelian case, 
$[\hat M_x,\hat\sigma_z]=0$,
the last term vanishes, and we get the same equation as previously. However, for the non-Abelian case, 
the last term introduces a coupling between the modes with $n_y$ and $-n_y$, 
and hence there is an explicit dependence on $n_y$. As a consequence of that, the degeneracy of the 
Landau levels is lifted.

For the particular case of $\hat M_x=\hat\sigma_y$, 
we get the following set of coupled equations 
($\epsilon=E-\hbar^2\kappa^2/2m$):
\begin{eqnarray}
\epsilon\vec u(n_y,x)&=&\left [ \frac{\hat p_x^2}{2m}+ 
\frac{B_0^2}{2m}\left ( x + \frac{2\pi\hbar n_y}{B_0 L}\right )^2 
\right ] \vec u(n_y,x) \nonumber \\
&+&\frac{\hbar\kappa}{m}\hat p_x\hat\sigma_y \vec u(-n_y,x)
\label{eps1} \\
\epsilon\vec u(-n_y,x)&=&\left [ \frac{\hat p_x^2}{2m}+ 
\frac{B_0^2}{2m}\left ( x - \frac{2\pi\hbar n_y}{B_0 L} \right )^2 
\right ] \vec u(-n_y,x) \nonumber  \\
&+&\frac{\hbar\kappa}{m}\hat p_x\hat\sigma_y \vec u(n_y,x).
\label{eps2}
\end{eqnarray}
The coupling prevents the re-absorption of $n_y$ in 
the definition of a new $q$ variable, as it was done in the Abelian case,
and hence the spectrum explicitely depends on $n_y$. 
Note that we are imposing periodic 
boundary conditions, and hence $x$ is in a ring of perimeter $L$. 
In this sense, $\pm L/2$ are the same point, and this must be taken 
into account when considering the harmonic oscillator 
potential in each equation. 

Note that the previous equations involve the coupling of harmonic oscillator 
wavefunctions centered in $\pm x_c (|n_y|)$, with 
$x_c (|n_y|)=2\pi\hbar |n_y| /B_0 L$. 
Hence, the 
smaller the overlapping between coupled wavefunctions, i.e. the large 
$x_c$, the smaller the coupling, and as a consequence 
only sufficiently small values of $n_y$ will be affected by the 
non-Abelian coupling. This point becomes 
clear after performing first order perturbation theory 
assuming a small coupling $\kappa$. A straightforward calculation shows 
that the lowest Landau levels, which correspond to the lowest eigenvalues 
of each harmonic oscillator, experience a maximal energy shift 
\begin{equation}
\frac{\Delta E}{\hbar\omega_c}=(\kappa l_c)\frac{n_y}{\Delta n_y}
e^{-n_y^2/\Delta n_y^2},
\end{equation}
where $l_c^2=\hbar/m\omega_c$ is the magnetic length, and 
$\Delta n_y=\sqrt{g/2\pi}$, with $g$ the degeneracy of the 
unperturbed Landau levels.
Note that for $n_y=0$ the first correction should be quadratic 
in $\kappa$, whereas for $n_y\neq 0$ it should be linear. 
Clearly, the relative importance of the non-Abelian corrections 
should decrease as $1/\sqrt{g}$. In particular, the 
maximal energy shift $\langle\Delta E\rangle$ averaged over the different $n_y$ 
can be approximated as 
$\langle \Delta E\rangle/\hbar\omega_c \simeq (\kappa l_c)/\sqrt{2\pi g}$.

\begin{figure}[ht]
\begin{center}
\psfrag{ny}{\footnotesize $n_y$}
\psfrag{E}{\footnotesize $E$}
\includegraphics[width=8 cm,angle=0]{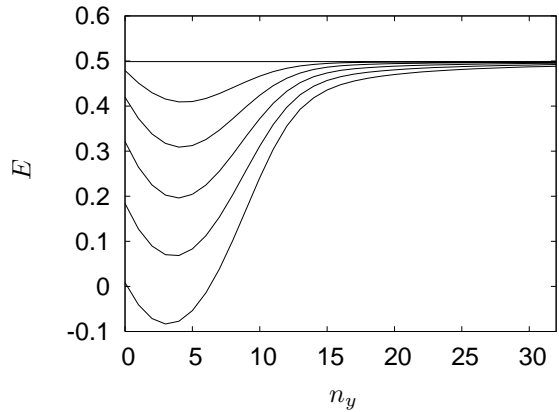}
\end{center}
\caption{Lowest eigenvalue $\epsilon/\hbar\omega_c$ 
as a function of $n_y$, for $g=128$ and 
$\kappa l_c=0, 0.2, 0.4, 0.6, 0.8, 1.0$ (from the 
uppermost to the lowermost curve).}
\label{fig:1}
\end{figure}

We have solved numerically for the eigenvalues of Eqs.~(\ref{eps1}) and 
(\ref{eps2}) imposing periodic boundary conditions, 
for different values of $g$ which controls the 
strength of the magnetic field applied, 
and $\kappa l_c$ which provides the strength of the non-Abelian 
corrections. The value of $L/l_c=\sqrt{2\pi g}$ is chosen in all simulations.
Fig.~\ref{fig:1} shows the behavior of the lowest 
eigenvalue as a function of $n_y$ 
for $g=128$ and $\kappa l_c=0, 0.2, 0.4, 0.6, 0.8, 1.0$ (from the 
uppermost to the lowermost curve). The figure follows approximately the 
perturbative result. For $n_y=0$ a higher order contribution appears, but 
note that a quadratic law follows for small $\kappa$ follows, and not a linear one, 
as in the case for $n_y\ne 0$. As expected from the previous calculations 
only values of $n_y$ up to the order of $\sqrt{g}$ 
contribute significantly to the shift of the lowest Landau level.

Figs. 2 show the behavior of the Landau levels for $g=128$, and 
$\kappa l_c=0$ (a), $0.6$ (b). The figures are presented as histograms 
in intervals of $0.05\hbar\omega_c$, in order to reveal more 
clearly the destruction of the Landau levels. Note that 
the gaps (of energy $\hbar\omega_c$) between the Landau levels 
are filled, and the peaks in the density of states are progressively reduced. 
For sufficiently large $\kappa$ the Landau level structure therefore disappears.

\begin{figure}[ht]
\begin{center}
\psfrag{degeneracy}{\footnotesize degeneracy}
\psfrag{E}{\footnotesize $E$}
\includegraphics[width=6 cm,angle=0]{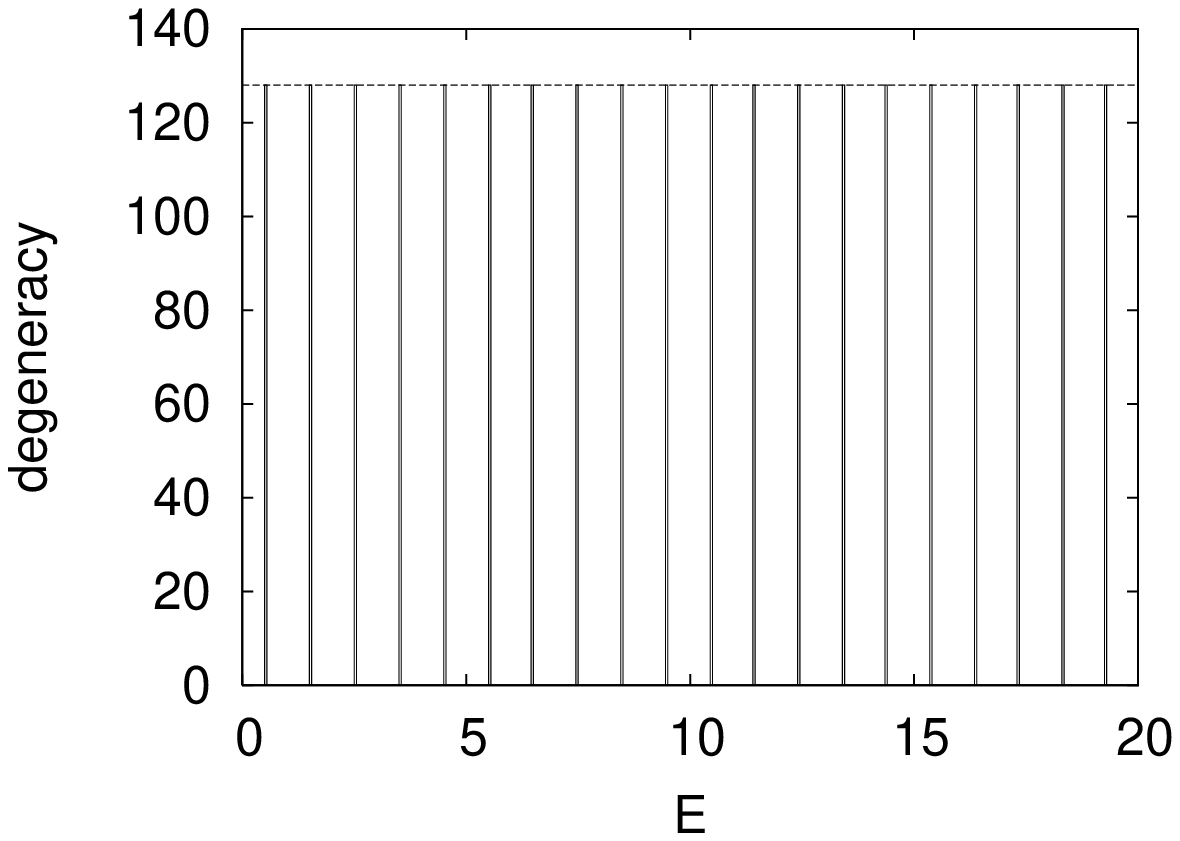}
\includegraphics[width=6 cm,angle=0]{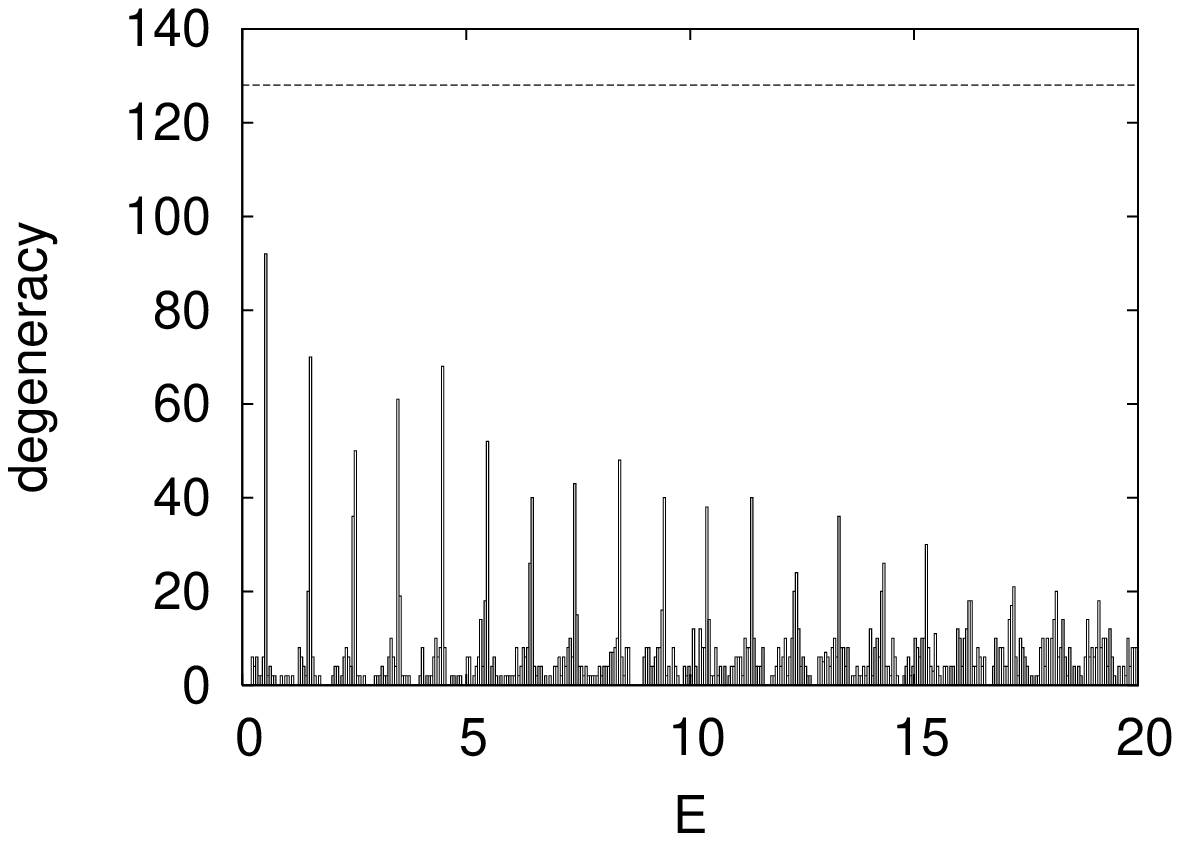}
\end{center}
\caption{Landau level structure for periodic boundary conditions, 
$g=128$, and $\kappa l_c=0$ (top) and $\kappa l_c=0.6$ (bottom). 
We employ (see text) $\hat M_x=\hat\sigma_y$ and $\hat M_y=\hat\sigma_z$.}
\label{fig:2}
\end{figure}

\subsection{Absorbing boundary conditions}

In the previous section we discussed how the non-Abelian 
character of the gauge field significantly modifies the text-book 
Landau level structure. In the following, we consider a 
slightly different physical scenario which is closer to the actual 
experimental conditions discussed in Sec.~\ref{sec:2}.
The particular procedure devised for the generation of the 
non-Abelian Landau gauge demands that the $x$ coordinate cannot be 
considered as periodic. We take the same box 
configuration as for the previous subsection, but assume 
absorbing boundary conditions in the $x$ direction, while 
keeping for simplicity periodic boundary conditions in the $y$-direction.
We consider exactly the same gauge discussed in the previous subsection. 
The spectrum is provided by Eqs.~(\ref{eps1}) and 
(\ref{eps2}) but imposing absorbing boundary conditions. 
Figs. 3 show the lowest Landau levels for the same cases discussed 
in Figs. 2. 

Even for the Abelian case the Landau level structure is 
of course affected by the absorbing boundary conditions. 
In the Abelian case, as discussed in the previous section, the 
problem reduces to two decoupled equations for harmonic oscillators centered 
at $\pm x_c (|n_y|)$. Clearly when $x_c$ approaches $L$ the 
levels of the resulting potential become largely distorted, leading to 
a significant modification of the Landau level structure when 
$n_y$ approaches $g$. This reduces the effective degeneracy of the 
lowest Landau levels to values smaller than $g$. The effective degeneracy, as 
shown in the figures, becomes smaller for higher Landau levels. 
The non-Abelian effect leads, as in the previous subsection, to 
the eventual destruction of the Landau level structure.

\begin{figure}[ht]
\begin{center}
\psfrag{degeneracy}{\footnotesize degeneracy}
\psfrag{E}{\footnotesize $E$ }
\includegraphics[width=6 cm,angle=0]{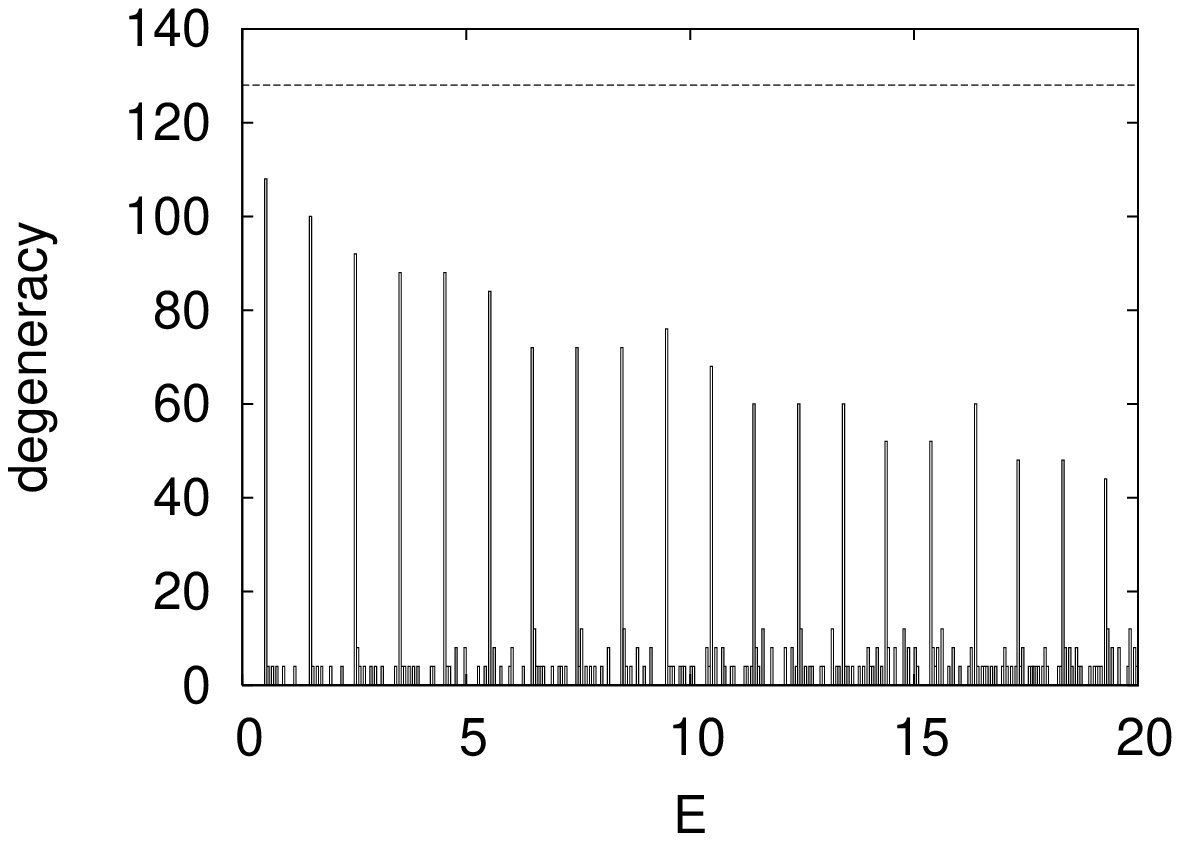}
\includegraphics[width=6 cm,angle=0]{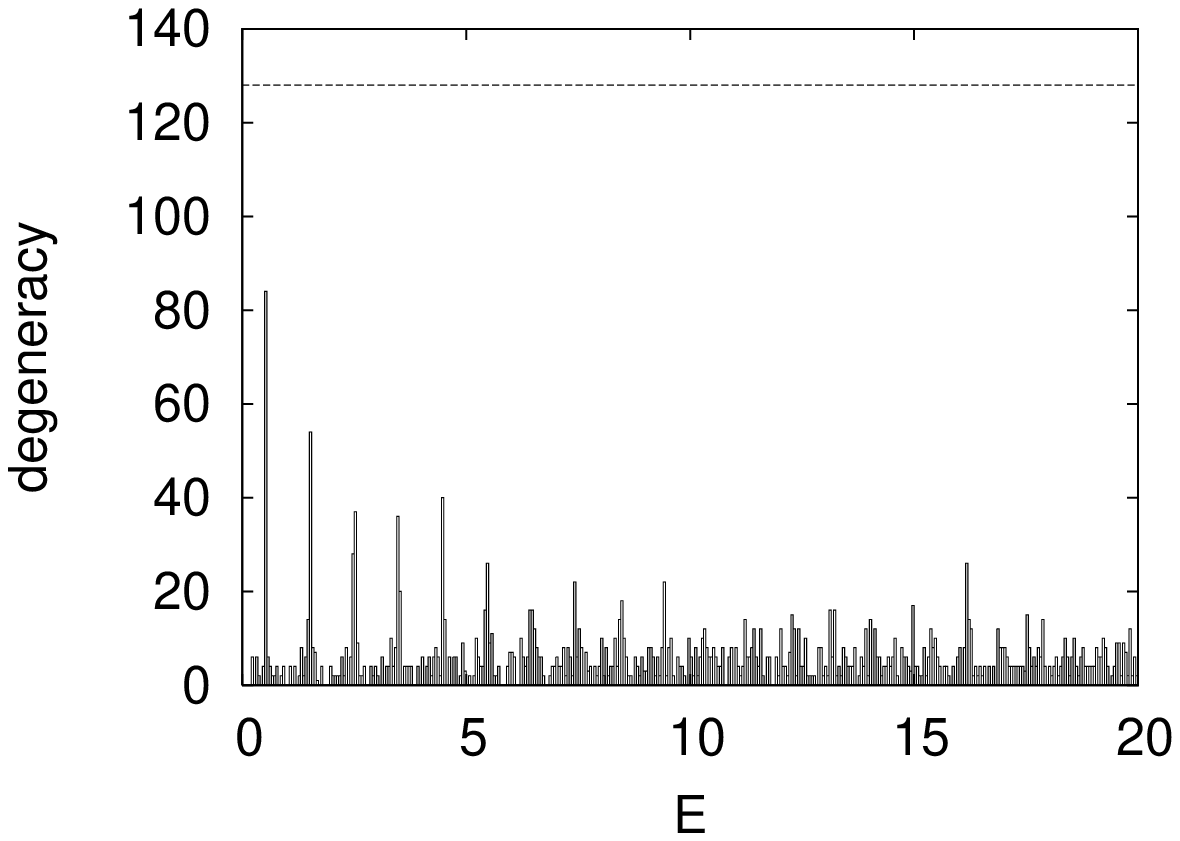}
\end{center}
\caption{Same cases considered in Fig.~\ref{fig:2} but with 
absorbing boundary conditions (see text).}
\label{fig:3}
\end{figure}

\subsection{Modified de Haas-van Alphen effect}

\begin{figure}[ht]
\begin{center}
\psfrag{B}{\footnotesize $B$}
\psfrag{chi}{\hspace{-0.7cm}\footnotesize $-d^2\bar E/dB_0^2$}
\includegraphics[width=7 cm,angle=0]{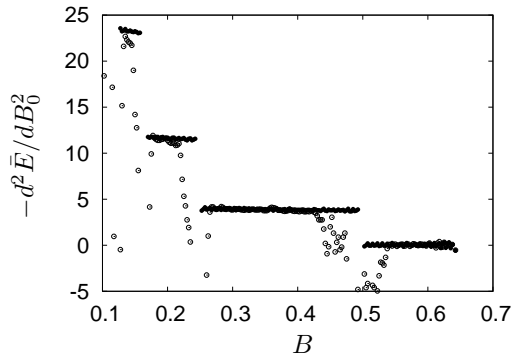}
\end{center}
\caption{Value of $-d^2\bar E/dB_0^2$ as a function 
of the applied magnetic field $B_0$, for the same case discuss in 
Fig.~\ref{fig:2}, and $\kappa=0$ (filled circles) 
and $\kappa=0.6$ (hollow circles).}
\label{fig:4}
\end{figure}

The destruction of the Landau level 
structure has experimentally relevant consequences for the 
behavior of cold atomic gases. As an example we can consider the 
case of an ideal two-component 
Fermi gas under the previously mentioned non-Abelian
gauge potential (we consider a temperature $T\ll T_F$, where $T_F$ 
is the Fermi temperature). Equivalently to the well-known 
de Haas-van Alphen effect \cite{dHvA}, we may study the 
energy per particle, $\bar E=E/N$, of the Fermi gas, as a function of the 
applied magnetic field $B_0$, or equivalently of $g$. This energy may be monitored 
by measuring the released energy in time-of-flight experiments. 
For $\kappa=0$ (Abelian case) $d^2\bar E(B_0)/dB_0^2$ presents 
a typical configuration of plateaux, due to the degeneracy of the 
Landau levels. The destruction of the Landau level structure significantly 
distorts this picture, rounding-off or eventually destroying this plateaux 
configuration (see Figs.~\ref{fig:4}).  
   
\section{Symmetric gauge}
\label{sec:5}

In this section we consider an ideal cold atomic sample in an isotropic 
harmonic trap of frequency $\omega$, in the presence of a
non-Abelian generalization of the symmetric gauge 
of the form $\hat A=\hat A_\rho \hat\rho + \rho \hat A_\varphi \hat\varphi$. 
Although the tripod scheme is not suitable for the experimental realization of this 
gauge, we include the analysis of this gauge field for completeness of our discussion. Other 
ways of generating non-Abelian gauge fields, as lattice techniques \cite{osterloh2005} 
should be employed in this case. In the following we 
consider $A_\rho=\hbar\kappa \hat U_\rho$, $A_\varphi=B_0 \hat U_\varphi$, where 
$\hat U_{\rho,\varphi}$ are linear combinations of $\{\hat{1},\hat\sigma_x,\hat\sigma_y,\hat\sigma_z \}$.

The corresponding time-independent Schr\"odinger equation is of the form
\begin{equation}
E\vec\psi=\frac{1}{2m} \left [-i\hbar\vec\nabla + \hat A \right ]^2\vec\psi +\frac{m\omega^2}{2}\rho^2\vec\psi.
\end{equation}
Performing the gauge transformation $\vec\psi=\exp [-i\hat A_\rho\rho/\hbar]\vec\phi$, the Schr\"odinger equation 
transforms into 
\begin{equation}
E\vec\psi=\frac{1}{2m} \left [-i\hbar\vec\nabla + \hat\varphi C_\varphi(\rho) \rho \right ]^2\vec\psi +\frac{m\omega^2}{2}\rho^2\vec\psi,
\end{equation}
where 
\begin{equation}
C_\varphi(\rho)=e^{i\hat A_\rho\rho/\hbar}\hat A_\varphi e^{-i\hat A_\rho\rho/\hbar}.
\end{equation}
Note that $C_\varphi$ becomes $\rho$ dependent and different from $\hat A_\varphi$ if $[\hat A_\rho,\hat A_\varphi]\neq 0$. 

If we now consider the solutions with angular momentum $l$, $\vec\phi=\vec R_l \rho^{|l|} e^{il\varphi}$,  we obtain 
\begin{eqnarray}
E\vec R_l &=& -\frac{1}{2} \left [ \frac{d^2}{d\rho^2}\vec R_l +\frac{ (2|l|+1) }{\rho} \frac{d}{d\rho}\vec R_l \right ] \nonumber \\
&+&\frac{1}{2}\left [ 1+C_\varphi(\rho)^2  \right ]\rho^2 \vec R_l 
+ l \hat C_\varphi (\rho)\vec R_l,
\end{eqnarray}
where we reduce the equations to a dimensionless form by employing oscillator units for the energy ($\hbar\omega$) and for the length 
($l_{ho}=\sqrt{\hbar/m\omega}$). In the previous equation
$\hat C_\varphi (\rho) \equiv (\omega_c/\omega) \exp [i\kappa \hat U_\rho \rho ] \hat U_\varphi  \exp [-i\kappa \hat U_\rho \rho]$, where $\omega_c=B_0/m$ is the corresponding cyclotron frequency. 

\begin{figure}[ht]
\begin{center}
\psfrag{E}{\footnotesize $E$}
\psfrag{B}{\footnotesize $B$}
\includegraphics[width=8 cm,angle=0]{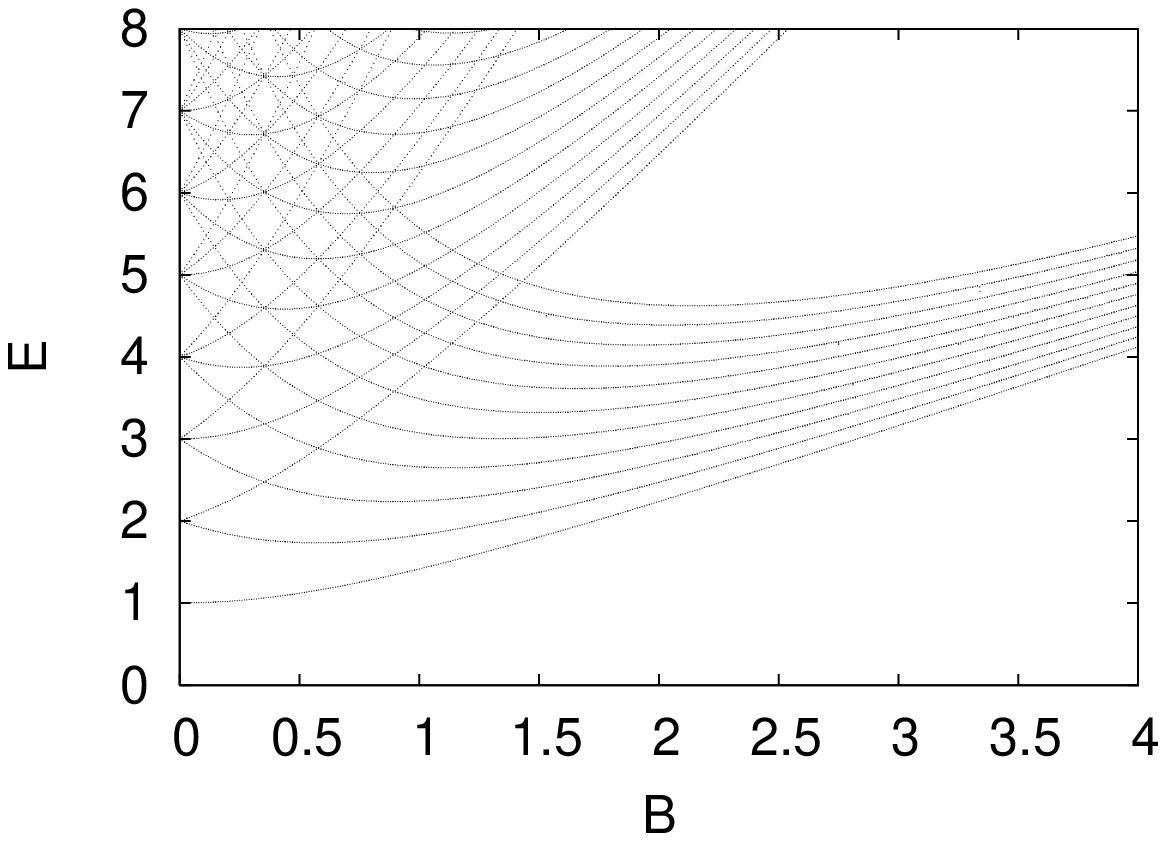}
\includegraphics[width=8 cm,angle=0]{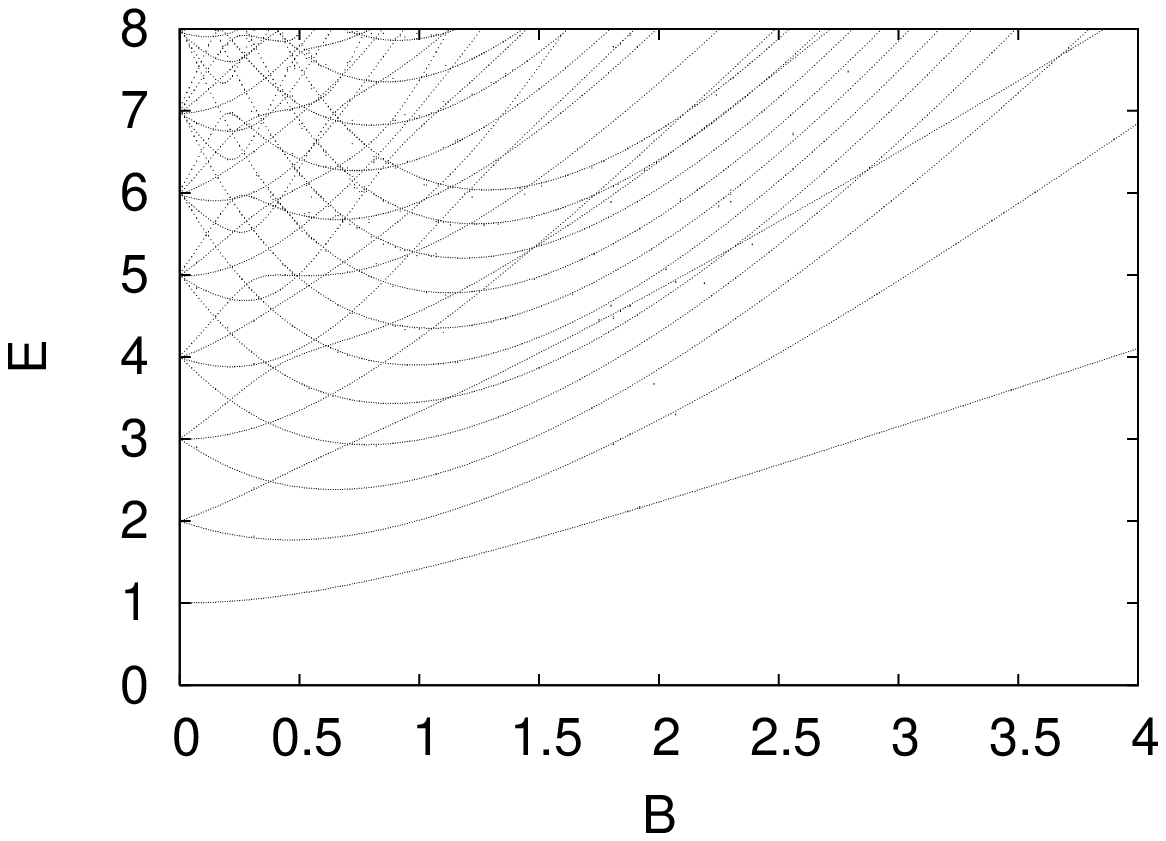}
\end{center}
\caption{
Fock-Darwin spectrum $E/\hbar\omega$ 
for the Abelian (top) and non-Abelian (bottom) 
cases discussed in the text 
as a function of $b_0=\omega_c/\omega$.}
\label{fig:5}
\end{figure}

\begin{figure}[ht]
\begin{center}
\psfrag{r}{\footnotesize $r$}
\psfrag{d}{\footnotesize $d$}
\includegraphics[width=6 cm,angle=0]
{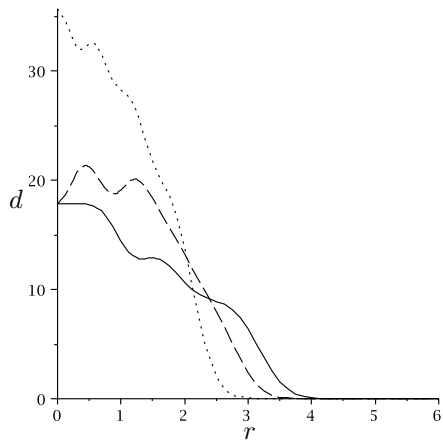}
\end{center}
\caption{Comparison between the density profile for an ideal Fermi gas 
occupying up to $56$ eigenlevels at zero temperature 
for the Abelian (solid) and non-Abelian cases discussed 
in the text with different values of $\kappa=1$ (dashed), and 
$\kappa=5$ (dotted).}
\label{fig:6}
\end{figure}

As mentioned above the non-Abelian character of the gauge field induces an additional $\rho$-dependent potential. It
severely distorts the standard Fock-Darwin spectrum which is expected for the Landau-level structure in the presence of 
a symmetric gauge and a harmonic potential, as it is shown in Fig.~\ref{fig:5}. An inspection of the level structure 
shows that not only the eigenenergies are modified, but also the ordering of the different eigenstates becomes distorted as a consequence 
of the non-Abelian potential. As a consequence of this extra $\rho$-dependent potential, an ideal Fermi gas at zero temperature shows 
a significantly distorted density profile in the presence of the non-Abelian gauge field, as shown in Fig.~\ref{fig:6}.

\section{Conclusions}
\label{sec:6}

In this paper we have analyzed the physics of ultracold gases in the presence of 
a non-Abelian gauge field. We have first studied how different types of non-Abelian 
fields may be created by means of relatively simple laser arrangements with atoms 
described by an electronic tripod level scheme, including a non-Abelian generalization of the 
Landau gauge. In a second part we have considered the non-trivial effects that the 
non-Abelian character has on the eigenlevel structure of the cold atomic system. In 
particular we have shown that exclusively due to the non-Abelian character of the field, the usual 
Landau level structure is severely distorted, and even eventually destroyed. We have shown that 
this effect may be observable in an equivalent experiment to the well-known de Haas-van Alphen 
effect. The distortion of the Landau levels leads to a significant modification of the 
usual plateaux-like signal characteristic for the de Haas-van Alphen effect. Finally, we have 
completed our analysis of a non-Abelian version of the symmetric gauge. We have shown that 
the Fock-Darwin spectrum is significantly distorted in the presence of non-Abelian fields, 
due to the presence of an extra potential, which is a purely non-Abelian effect. 

\ack
This work was supported by the Deutsche Forschungsgemeinschaft 
(SFB-TR21, SFB407, SPP1116), the European Graduate College
``Interference and Quantum Applications'', 
and the UK Engineering and Physical Sciences Research Council.

\end{document}